\documentclass[lettersize,journal]{IEEEtran}
\usepackage{amsmath,amsfonts}
\usepackage{algorithmic}
\usepackage{algorithm}
\usepackage{array}
\usepackage{pifont}
\usepackage[caption=false,font=normalsize,labelfont=sf,textfont=sf]{subfig}
\usepackage{fontawesome5}
\usepackage{textcomp}
\usepackage{stfloats}
\usepackage{url}
\usepackage{verbatim}
\usepackage{graphicx}
\usepackage{caption}
\usepackage[section]{placeins}
\usepackage{tabularray}
\usepackage{adjustbox} 
\usepackage{float}
\usepackage{color}
\usepackage{booktabs}
\usepackage{hyperref} 

\usepackage{booktabs}
\usepackage{colortbl}

\usepackage{ulem}
\usepackage[usenames,svgnames,table]{xcolor}
\usepackage{hyperref}
\usepackage{url}
\definecolor{darkblue}{rgb}{0, 0, 0.5}
\definecolor{beaublue}{rgb}{0.74, 0.83, 0.9}
\definecolor{gainsboro}{rgb}{0.86, 0.86, 0.86}
\definecolor{kleinblue}{rgb}{0,0.18,0.65}
\hypersetup{colorlinks=true,citecolor=kleinblue, linkcolor=kleinblue, urlcolor=kleinblue}

\begin{document}

\title{Personality Analysis from Online Short Video Platforms with Multi-domain Adaptation}

\author{Sixu An\textsuperscript{$*$}, Xiangguo Sun\textsuperscript{$*$, \faEnvelope[regular]}, Yicong Li, Yu Yang, Guandong Xu
\thanks{Sixu An, Yicong Li, Yu Yang, Guandong Xu are with The Education University of Hong Kong, Hong Kong SAR, s1154340@s.eduhk.hk, liyicong123@outlook.com, \{yangyy, gdxu\}@eduhk.hk, Xiangguo Sun contributed equally to the first author and is the corresponding author, with The Chinese University of Hong Kong, Hong Kong SAR, xiangguosun@cuhk.edu.hk}}

% The paper headers
\markboth{Journal of \LaTeX\ Class Files,~Vol.~14, No.~8, August~2021}%
{Shell \MakeLowercase{\textit{et al.}}: A Sample Article Using IEEEtran.cls for IEEE Journals}

\maketitle

\begin{abstract}
Personality analysis from online short videos has gained prominence due to its applications in personalized recommendation systems, sentiment analysis, and human-computer interaction. Traditional assessment methods, such as questionnaires based on the Big Five Personality Framework, are limited by self-report biases and are impractical for large-scale or real-time analysis. Leveraging the rich, multi-modal data present in short videos offers a promising alternative for more accurate personality inference. However, integrating these diverse and asynchronous modalities poses significant challenges, particularly in aligning time-varying data and ensuring models generalize well to new domains with limited labeled data. In this paper, we propose a novel multi-modal personality analysis framework that addresses these challenges by synchronizing and integrating features from multiple modalities and enhancing model generalization through domain adaptation. We introduce a timestamp-based modality alignment mechanism that synchronizes data based on spoken word timestamps, ensuring accurate correspondence across modalities and facilitating effective feature integration. To capture temporal dependencies and inter-modal interactions, we employ Bidirectional Long Short-Term Memory networks and self-attention mechanisms, allowing the model to focus on the most informative features for personality prediction. Furthermore, we develop a gradient-based domain adaptation method that transfers knowledge from multiple source domains to improve performance in target domains with scarce labeled data. Extensive experiments on real-world datasets demonstrate that our framework significantly outperforms existing methods in personality prediction tasks, highlighting its effectiveness in capturing complex behavioral cues and robustness in adapting to new domains. We open our datasets and code at \url{https://github.com/Anne6645/personality_analysis}

\end{abstract}


\section{Introduction}
\label{sec:introduction}

\IEEEPARstart{P}{ersonality} analysis has long been a central topic in psychological science and has gained increasing importance in recent years due to its wide-ranging applications. It plays a crucial role in various domains such as personalized recommendation systems \cite{ning2019personet,yang2022personality}, sentiment analysis \cite{zhang2024modeling, liu2020context}, and human-computer interaction. Accurately identifying an individual's personality can enable tailored experiences and services, enhancing user satisfaction and engagement. However, personality traits are inherently latent characteristics that are not directly observable, making the assessment of personality a challenging task.

\begin{figure}[t]
\centering
\includegraphics[width=0.38\textwidth]{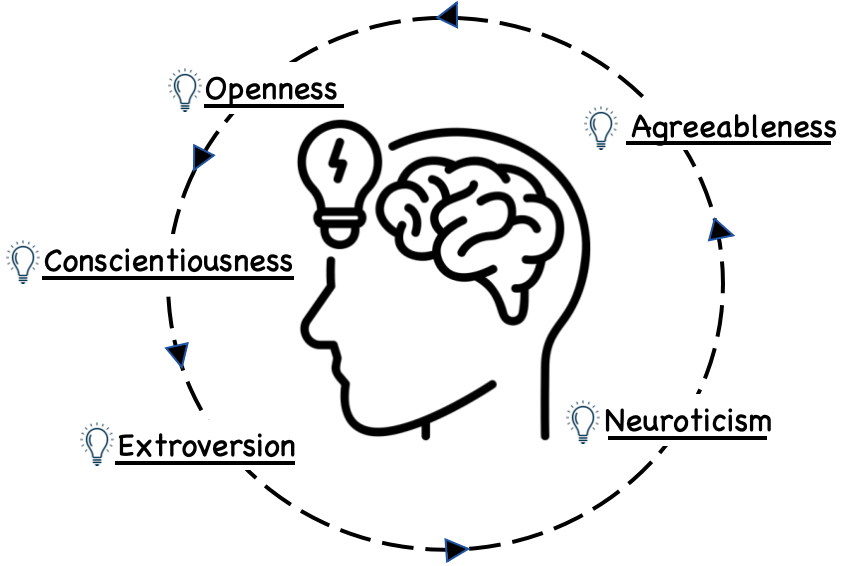}
\caption{Big Five Personality Traits}\vspace{-2em}
\label{fig:big5}
\end{figure}

Traditionally, psychologists have employed structured methods to evaluate an individual's personality. One of the most widely accepted models is the \textit{Big Five Personality Traits} (as shown in Figure \ref{fig:big5}), which assesses personality across five key dimensions: openness, conscientiousness, extraversion, agreeableness, and neuroticism. To determine an individual's position on these dimensions, conventional approaches often rely on well-designed questionnaires and psychological inventories that analyze self-reported responses. While these methods are grounded in rigorous psychometric principles, they have notable limitations. Self-reported data can be influenced by social desirability bias, where respondents tailor their answers to be viewed favorably. Additionally, administering and processing these surveys can be time-consuming and resource-intensive, making them less practical for large-scale or real-time applications.

With the advent of online video social platforms like TikTok\footnote{\url{https://www.tiktok.com/}} and others, there is a growing opportunity to analyze personality traits through digital means. Users increasingly share selfie videos online, providing a wealth of data that captures not only their visual appearance but also their speech patterns, facial expressions, and environmental context. Compared to static questionnaires, these multi-modal data offer richer insights into an individual's intrinsic traits. Unlike traditional social media platforms that primarily feature text or images, video platforms enable the observation of dynamic behaviors and interactions, which are crucial for understanding personality. This shift opens up new possibilities for applications such as online job interviews, remote education, and personalized content delivery, where assessing personality from videos can significantly enhance outcomes.

Recent research has begun to explore the potential of analyzing personality traits through online media instead of traditional surveys. Behavioral observations from personal photographs \cite{segalin2017your, liu2016analyzing, celli2014automatic} and short videos \cite{sun2022your} have been utilized to glean personality insights. For instance, \cite{segalin2017your} analyzed Facebook profile pictures to infer personality traits, while \cite{liu2016analyzing} and \cite{celli2014automatic} leveraged social media images for similar purposes. However, photograph-based approaches have limitations, as individuals often curate their online images, sharing selective moments that may not accurately represent their typical behaviors or personality, leading to biased data and potentially inaccurate predictions. \textit{\dotuline{In contrast, short videos provide a more comprehensive medium for personality analysis.}} They capture changes in facial expressions, body movements, speech patterns, and contextual scenes—all of which are significant indicators in psychological assessments of personality. Recognizing this, researchers have started to model the audio, visual, and textual features present in short videos \cite{zhang2016deep, wei2017deep, sarkar2014feature}. For example, \cite{zhang2016deep} developed a Deep Bimodal Regression model combining audio and visual modalities to predict scores on the Big Five personality traits. Similarly, \cite{wei2017deep} employed convolutional neural networks to extract visual features and linear regression for audio features, while \cite{sarkar2014feature} conducted an in-depth analysis using logistic regression on audio, video, and text features.

\begin{figure}[t]
\centering
\includegraphics[width=0.45\textwidth]{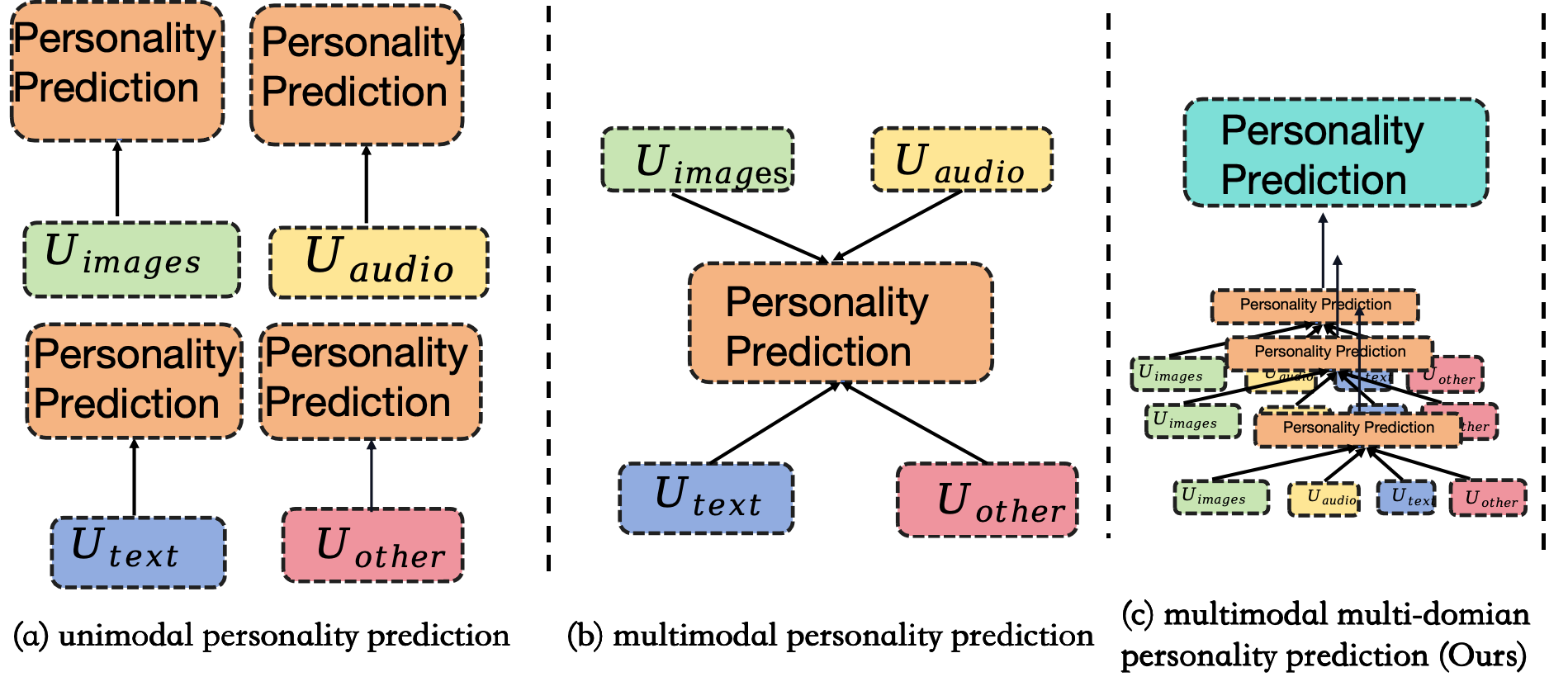}
\caption{Comparison of Existing Works}\vspace{-2em}
\label{fig:intro}
\end{figure}

Despite these advances, existing multi-modal personality prediction methods often rely on large volumes of high-quality short videos with high-resolution visuals and clear audio to achieve satisfactory performance. Moreover, many approaches depend heavily on supervised learning techniques that require extensive labeled datasets. Collecting and annotating such multi-modal data is both expensive and time-consuming. Manual annotation introduces the potential for subjectivity and inconsistency, which can affect the reliability of the analysis. Consequently, detecting personality traits from online video platforms presents significant challenges, particularly in the following areas:

The \uline{\textit{\textbf{First Challenge}}} is identifying the most important features from multiple modalities to optimize the use of a limited number of short videos for accurate personality analysis. The complexity arises from the need to effectively integrate diverse data types—visual cues, auditory signals, textual content, and contextual information—each contributing uniquely to personality inference. Existing methods may not adequately address the alignment and synchronization of these modalities, leading to fragmented or incomplete representations that hinder predictive accuracy.

The \uline{\textit{\textbf{Second Challenge}}} is effectively utilizing a small number of high-quality short videos to achieve strong generalization in personality analysis. Models trained on specific datasets may struggle to generalize across different domains due to variations in cultural contexts, linguistic expressions, and recording conditions. The scarcity of labeled data in new or underrepresented domains exacerbates this issue, limiting the applicability of the models in real-world scenarios where data diversity is the norm.

In this paper, we propose an effective multi-modal personality analysis framework designed to overcome these challenges. \textbf{To address the first challenge}, we introduce a semantic unit method for feature extraction and alignment, which synchronizes multi-modal data based on spoken words. This ensures that features from different modalities correspond accurately at each moment in the video, facilitating effective integration. Within this module, we employ self-attention mechanisms to discern the significance of features across various modalities. By assigning weights to features based on their relevance to personality prediction, the model focuses on the most informative aspects of the data, enhancing analytical accuracy. \textbf{To tackle the second challenge}, we propose a multi-domain adaptation method that transfers domain knowledge across multiple domains to alleviate the data sparsity problem. This approach leverages information from data-rich source domains to enhance learning in data-scarce target domains. By computing gradient similarities between source and target domains, our model adapts to emphasize learning from source domains that are most relevant to the target domain. This method improves the model's generalization capabilities, enabling more accurate predictions even when limited data is available in certain domains.

Our main contributions are summarized as follows:

\begin{itemize}
    \item We propose an effective multi-modal personality analysis framework that effectively integrates facial expressions, audio signals, textual content, and background information from short videos for personality prediction.
    \item We introduce a semantic unit modality alignment mechanism that synchronizes multi-modal data based on spoken word timestamps, ensuring accurate correspondence across modalities and enhancing feature representation.
    \item We develop a gradient-based domain adaptation method that transfers knowledge from multiple source domains to target domains with limited labeled data, enhancing model generalization and performance in few-shot learning scenarios.
    \item We validate the effectiveness of our proposed framework through extensive experiments on real-world datasets, demonstrating significant improvements over existing methods in personality prediction tasks.
\end{itemize}

By addressing both the feature integration and domain adaptation challenges, our framework advances in personality analysis from online short videos. The rest of the paper is organized as follows. Section \ref{sec:related_work} introduces the related work of personality analysis and domain adaption. The problem definition is in Section \ref{sec:problem_definition}. Section \ref{sec:method} elucidates the detailed methodology, and Section \ref{sec:experiment} presents the results of the experiments and the analysis. The last section is the conclusion of the paper. 

\section{Related Work}\label{sec:related_work}

Personality analysis has garnered significant interest in recent years due to its applications in psychology, human-computer interaction, and personalized services. Researchers have explored various data sources and methodologies to predict personality traits effectively. Early studies focused on analyzing behavioral patterns from personal devices. For instance, \cite{schoedel2019digital} examined communication habits and app usage data collected from Android smartphones to infer personality characteristics. Similarly, \cite{rastogi2020intelligent} leveraged data from Internet of Things (IoT) devices for personality assessment, highlighting the potential of ubiquitous computing in this domain.

With the rise of social media platforms, user-generated content has become a rich resource for personality prediction. \cite{philip2019machine} utilized user activities and interactions across various social media sites to analyze personality traits, demonstrating the value of digital footprints in psychological profiling. In these early computational approaches, traditional machine learning algorithms such as Naive Bayes (NB), Support Vector Machines (SVM), and XGBoost classifiers were commonly employed \cite{nisha2022comparative}. As the field progressed, advanced techniques from computer vision and natural language processing \cite{sun2020group} were adopted to enhance predictive performance. For example, \cite{maliki2020personality} analyzed handwriting signatures to gain insights into personality traits, while \cite{sajeevan2021detection} also utilized handwriting data for personality assessment. In the context of social media, \cite{kamalesh2022personality} proposed a Binary-Partitioning Transformer (BPT) model combined with Term Frequency and Inverse Gravity Moment features to predict personality traits using Twitter data.

The evolution of personality analysis techniques has led to their application in various fields such as recommender systems, recruitment processes, and sentiment analysis. For instance, \cite{sudha2021personality} evaluated interviewees' personalities through quizzes and curriculum vitae assessments to aid in hiring decisions. Additionally, \cite{escalera2018guest} introduced an automatic framework for real-time personality analysis, emphasizing the practical utility of these methods. Recent advancements in artificial intelligence, particularly the emergence of Large Language Models (LLMs), have further expanded the capabilities of personality analysis. In 2023, \cite{wen2023desprompt} explored the use of personality-descriptive prompts to fine-tune pre-trained language models for few-shot personality recognition, showcasing the potential of LLMs in this domain.

Despite these advancements, there remains a notable gap in the literature regarding video-based personality analysis, which inherently involves multi-modal information such as text, facial expressions, and vocal cues. Existing multi-modal approaches typically extract features from text, facial images, and voice recordings separately but often lack effective methods for aligning all modalities and preserving the continuity of time-series information. This misalignment can hinder the model's ability to accurately capture the dynamic interplay between different types of data.

In our work, we address this gap by proposing a timestamp-based modality alignment technique that synchronizes different modalities based on assigned timestamps. This method aligns four modalities—textual content, facial expressions, audio signals, and background information—ensuring that the temporal dynamics of the video are preserved. By maintaining the consistency and accuracy of the information conveyed across modalities, our approach enhances the reliability of personality prediction from online short videos. This alignment not only improves the integration of multi-modal data but also leverages the sequential nature of videos to capture nuanced personality cues that may be missed by methods lacking temporal synchronization.
\section{Preliminary}\label{sec:problem_definition}
In this section, we first give a background about the Big Five Traits, then introduce our motivations, and finally formalize our objectives.

\subsection{Background of the Big Five Personality Traits}

Personality analysis aims to identify and quantify individual differences in behavioral patterns, thought processes, and emotional responses. Among the various models developed to describe personality, the \textit{Big Five Personality Traits} model has emerged as a widely accepted framework due to its robust empirical support and comprehensive nature. This model delineates personality along five fundamental dimensions: \textit{Openness to Experience} ($ O $), \textit{Conscientiousness} ($ C $), \textit{Extraversion} ($ E $), \textit{Agreeableness} ($ A $), and \textit{Neuroticism} ($ N $). Specifically, Openness to Experience reflects imagination, creativity, curiosity, and a preference for novelty and variety. Conscientiousness indicates self-discipline, organization, dependability, and a goal-oriented approach. Extraversion characterizes sociability, assertiveness, energy levels, and a tendency toward positive emotions. Agreeableness involves compassion, cooperation, trust, and a concern for social harmony. Neuroticism relates to emotional instability, anxiety, moodiness, and susceptibility to negative emotions. Mathematically, an individual's personality profile can be represented as a vector $ \mathbf{p} \in \mathbb{R}^5 $:
\begin{equation}
\mathbf{p} = [ p_O, p_C, p_E, p_A, p_N ],
\end{equation}
where $ p_O, p_C, p_E, p_A, p_N $ are scalar values quantifying the levels of openness, conscientiousness, extraversion, agreeableness, and neuroticism, respectively. Each trait is typically measured on a continuous scale, often ranging from low to high scores. Compared to other personality models such as the Myers-Briggs Type Indicator (MBTI) or the Enneagram, the Big Five model stands out for its dimensional approach and strong empirical validation across cultures and populations. It provides a holistic methodology for personality assessment, capturing a wide spectrum of human behavior and traits.

\subsection{Motivations}

\ding{168} \textit{\dotuline{Why Multi-Modal Learning for Personality Analysis?}}
Personality manifests through a rich tapestry of behavioral cues expressed across multiple modalities, and relying on a single modality may overlook critical aspects of these expressions, thereby limiting the accuracy of personality assessments. \textbf{Multi-modal learning} enables the integration of diverse data sources—such as facial expressions, vocal characteristics, linguistic content, and background context—to capture the complexity of personality traits more comprehensively. For instance, facial features can reveal underlying emotions not explicitly communicated through words, while variations in speech patterns—including tone, pitch, volume, and speaking rate—are indicative of certain personality traits. The choice of words and linguistic style provides direct indications of thought processes, and the context or setting in which a person appears offers clues about their preferences, lifestyle, and social context. Short videos serve as an ideal medium to capture these multi-modal signals simultaneously, providing a sequential stream of visual and auditory information that reflects the dynamic interplay of these cues. By extracting and integrating features from these modalities—facial features (\( X_{\text{face}} \)), audio features (\( X_{\text{audio}} \)), textual content (\( X_{\text{text}} \)), and background information (\( X_{\text{background}} \))—we can develop a more robust and accurate model for personality analysis, capturing the full spectrum of behavioral indicators and leading to more reliable and nuanced predictions.

\ding{168} \textit{\dotuline{Why Domain Adaptation for Personality Analysis?}} However, personality detection models often face challenges when applied across different domains due to variations in cultural expressions, language usage, video styles, and environmental contexts. A model trained on data from one domain may not generalize well to another, resulting in decreased performance. Collecting and annotating large amounts of labeled data for every new target domain is impractical, making \textbf{domain adaptation} a crucial technique. Domain adaptation aims to transfer knowledge from one or more source domains to a target domain with limited labeled data, thereby improving model performance without the need for extensive data collection in the new domain. By incorporating domain adaptation techniques, we enhance the robustness and applicability of our personality detection model across diverse settings, ensuring that it remains effective even when applied to new domains with minimal additional data. This approach addresses practical challenges such as domain shift, data scarcity, and variability in expressions, ultimately leading to more accurate and generalizable personality assessments.

\subsection{Objective Formulation}

Our objective is to develop a domain-adaptive, multi-modal personality prediction model that accurately estimates the Big Five personality traits vector $ \mathbf{p} $ for individuals based on short video data. The problem can be formally defined as follows.

Let $ \mathcal{D}_s = \{ (\mathbf{x}_i^s, \mathbf{p}_i^s) \}_{i=1}^{N_s} $ denote the source domain data, where $ \mathbf{x}_i^s $ represents the multi-modal features extracted from the $ i $-th video in the source domains, and $ \mathbf{p}_i^s $ is the corresponding personality vector. The target domain data is given by $ \mathcal{D}_t = \{ (\mathbf{x}_j^t, \mathbf{p}_j^t) \}_{j=1}^{N_t} $, where $ N_t \ll N_s $, indicating that the target domain has limited labeled data. We aim to learn a personality prediction model $ f_\theta : \mathbf{x} \rightarrow \mathbf{p} $, parameterized by $ \theta $, that minimizes the prediction error on the target domain while effectively utilizing the source domain data. The optimization objective can be expressed as:
\begin{equation}
\min_{\theta} \mathcal{L}_{\text{total}}(\theta) = \mathcal{L}_{\text{target}}(\theta) + \lambda \mathcal{L}_{\text{adaptation}}(\theta),
\end{equation}
where $ \mathcal{L}_{\text{target}}(\theta) $ is the loss on the target domain, $ \mathcal{L}_{\text{adaptation}}(\theta) $ is the domain adaptation loss, and $ \lambda $ is a hyperparameter that balances the two components. The target domain loss is defined as:
\begin{equation}
\mathcal{L}_{\text{target}}(\theta) = \frac{1}{N_t} \sum_{j=1}^{N_t} \mathcal{L}\left( f_\theta(\mathbf{x}_j^t), \mathbf{p}_j^t \right),
\end{equation}
Where $ \mathcal{L} $ is the Mean Squared Error (MSE) between the predicted and true personality trait vectors:
\begin{equation}
\mathcal{L}\left( f_\theta(\mathbf{x}), \mathbf{p} \right) = \frac{1}{5} \sum_{k=1}^{5} \left( f_\theta(\mathbf{x})_k - p_k \right)^2,
\end{equation}
with $ f_\theta(\mathbf{x})_k $ being the predicted score for the $ k $-th personality trait.

The domain adaptation loss $ \mathcal{L}_{\text{adaptation}}(\theta) $ aims to minimize the discrepancy between the source and target domains. This can be implemented using techniques such as gradient reversal layers, adversarial training, or gradient-based similarity measures, which align the feature distributions or adapt the model parameters to reduce domain shift.

\section{Personality Detection based on Multi-Modal Short Video Analysis}\label{sec:method}

\subsection{Overview of the Framework}

We introduce a multi-modal video personality prediction framework designed to capture and integrate diverse behavioral cues from short videos for accurate personality trait prediction. As shown in Figure \ref{fig:framework}, the framework comprises three key modules: (1) multi-modal feature extraction and alignment, (2) personality detection based on multi-modal analysis, and (3) enhancement of personality analysis via domain adaptation. By systematically integrating these components, our framework leverages rich information from facial expressions, background context, audio signals, and textual content while ensuring robustness and generalizability across different domains.

The motivation behind this pipeline is to address the challenges inherent in personality prediction from online short videos, including handling heterogeneous multi-modal data, aligning asynchronous modalities, and adapting models to new domains with limited labeled data. The framework first extracts and aligns features from the four modalities using a Timestamp Modality Alignment mechanism based on spoken word timestamps. It then processes these features through temporal modeling and cross-modal fusion techniques to capture both intra- and inter-modal dynamics. Finally, a gradient-based domain adaptation method enhances model performance on target domains with scarce data by leveraging similarities between source and target domains. This integrated approach allows the model to effectively capture complex behavioral cues and remain robust across diverse settings.

\subsection{Multi-Modal Features for Video Analysis}

In this research, we decompose short videos $ V $ into four data modalities: speakers' facial expressions, background information, audio signals, and textualized speech content. The selection of these modalities is motivated by their collective ability to capture comprehensive cues related to personality traits. Specifically, non-verbal cues such as facial movements and expressions can reveal underlying emotions and personality characteristics; the environment or context in which a person appears can provide insights into their lifestyle, preferences, and social context; elements like tone, pitch, and speech rhythm convey paralinguistic information essential for understanding affective states and personality; and the semantic content and linguistic patterns of speech offer direct indications of thought processes and personality traits.

However, a key challenge arises because these modalities are time-varying and have different granularities. Text is typically segmented into words, images into frames, and audio into time intervals, making alignment difficult. Since the modalities $ \{ X_{\text{face}}, X_{\text{background}}, X_{\text{audio}}, X_{\text{text}} \} $ are captured independently, they inherently lack synchronization. To address this, we observe that during speech in a video, each timestamp corresponds to a spoken word, associated facial expression, audio signal, and background scene. This one-to-one correspondence allows us to align the modalities effectively.

\subsubsection*{\ding{182}  \textbf{Timestamp Modality Alignment (TMA)}} We propose a timestamp modal alignment (TMA) mechanism to synchronize the four modalities at each timestamp based on minimal semantic units—specifically, individual spoken words. This alignment ensures that multi-modal features correspond precisely, facilitating a comprehensive analysis of personality traits. To implement this alignment, we segment the modalities using the timestamps of words as they appear in the audio stream. These timestamps are obtained using the Connectionist Temporal Classification (CTC) algorithm \cite{kurzinger2020ctc}, applied via the pre-trained Wav2Vec2 model \cite{baevski2020wav2vec}. By aligning the audio segments with the spoken words, we establish a temporal framework for synchronization. Similarly, we slice the corresponding video frame sequences according to these timestamp segments. This process ensures that the visual (both facial expressions and background), audio, and textual data for each segment are matched accurately. 

Formally, given a short video $ v \in V $, we denote its transcribed speech content as $ v_{\text{text}} = \{ w_1, w_2, \ldots, w_n \} $, where $ w_i $ is the $ i $-th word spoken, associated with timestamp $ \tau_i $. At each timestamp $ \tau_i $, we have: (1) Frame Extraction. We extract the video frames corresponding to $ \tau_i $, resulting in $ f_{\tau_i} $; (2) Audio Segment. We obtain the audio segment $ v_{\text{audio}, \tau_i} $ corresponding to the duration of $ w_i $; (3) Facial Image. We extract the facial region $ v_{\text{face}, \tau_i} $ from $ f_{\tau_i} $; (4) Background Information. We capture the background $ v_{\text{background}, \tau_i} $ from $ f_{\tau_i} $; (5) Word Token. We have the word $ w_{\tau_i} = w_i $ spoken at $ \tau_i $. This approach results in a data tuple for each timestamp $ \tau_i $ with four modal features like $ D_{\tau_i} = \{ v_{\text{face}, \tau_i}, v_{\text{background}, \tau_i}, v_{\text{audio}, \tau_i}, w_{\tau_i} \} $. To handle cases where a person is not speaking (leading to missing $ D_{\tau_i} $) and to prevent data imbalance due to varying speech lengths, we standardize the text length. We set $ n = 70 $ words for each video's text data (an empirical length for most short videos online). Videos with fewer than 70 words are padded with a special token $[ \text{UNK} ]$. Thus, we obtain a consistent sequence $ V = \{ D_{\tau_1}, D_{\tau_2}, \ldots, D_{\tau_n} \} $ for each video.

\begin{figure*}[t]
\centering
\includegraphics[width=0.7\textwidth]{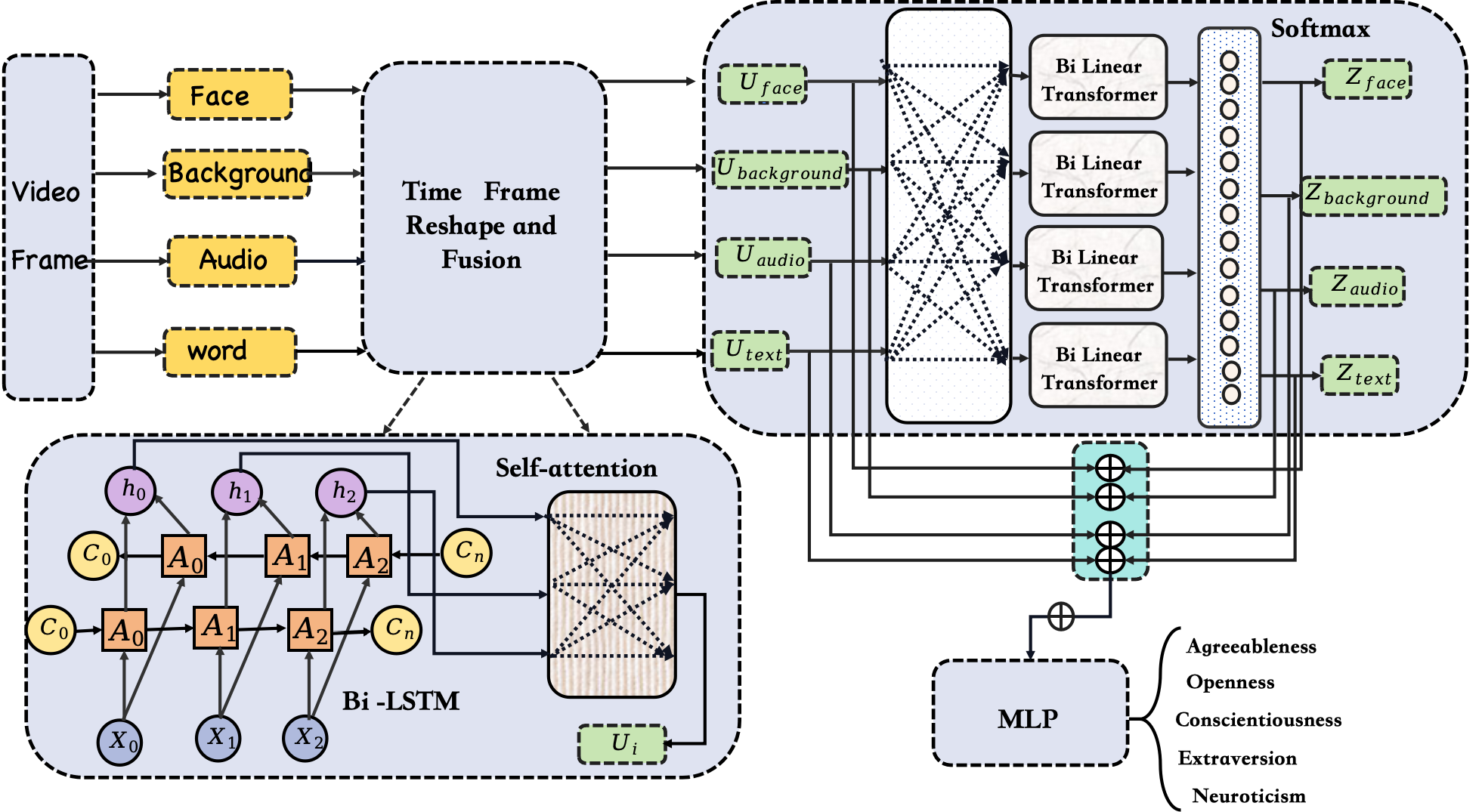}
\caption{The Flowchart of Multi-model personality prediction.}
\label{fig:framework}
\end{figure*}

\subsubsection*{\ding{183}  \textbf{Modal Feature Extraction}} After aligning the data, we extract features from each modality in $ D_{\tau_i} $ using specialized pre-trained models, ensuring that the representation of each modality captures both low-level and high-level information.

 \ding{168} \textit{\dotuline{Visual Features:}} The visual modality consists of both facial expressions and background context. These two visual features are combined to form the overall visual representation $ v_{\text{visual}, \tau_i} $ for each timestamp $ \tau_i $.:
\begin{itemize}
    \item Facial Features: Recognizing that facial expressions can reflect underlying personality traits, we utilize an effective facial feature extractor based on the Face Recognition library\footnote{\url{https://github.com/ageitgey/face_recognition}}. This tool extracts facial features from $ v_{\text{face}, \tau_i} $, providing a detailed representation of the speaker's facial characteristics.
    \item Background Features: The environment in which the speaker appears can influence first impressions and provide context about their personality. To capture this, we employ TimeSformer \cite{bertasius2021space}, a pre-trained video model known for its excellent performance in extracting high-level features from video frames. TimeSformer processes $ v_{\text{background}, \tau_i} $, producing a feature vector that represents the global background context.
\end{itemize}

\ding{168} \textit{\dotuline{Audio Features:}} Auditory cues such as tone, speed, and pitch can vary with personality. By concatenating the MFCCs and the features from WavLM, we obtain a comprehensive audio representation $ v_{\text{audio}, \tau_i} $ for each timestamp. To capture these nuances:
\begin{itemize}
    \item Low-Level Audio Features: We compute the Mel Frequency Cepstral Coefficients (MFCCs) of the audio segment $ v_{\text{audio}, \tau_i} $. MFCCs simulate how humans perceive sound and provide a compact representation of the audio signal's spectral properties.
    \item High-Level Audio Features: We use WavLM \cite{chen2022wavlm}, a pre-trained model for speech representation learning, to extract high-level features from $ v_{\text{audio}, \tau_i} $.
\end{itemize}

\ding{168} \textit{\dotuline{Textual Features:}} For the textual modality, we transform the word token $ w_{\tau_i} $ into an embedding using the token embedding layer of the language model employed in our framework. This approach ensures that the textual data is represented in the same embedding space as the language model, facilitating downstream processing.

Note that we do not fine-tune these pre-trained models for feature extraction during our training process; they serve solely to provide rich, pre-processed representations. After processing each modality, for the $ i $-th video, we obtain sequences of representation vectors as follows: visual $ v_i \in \mathbb{R}^{d_v \times n} $, audio $ a_i \in \mathbb{R}^{d_a \times n} $, and textual $ t_i \in \mathbb{R}^{d_t \times n} $ where $ n $ is the sequence length (standardized to 70), and $ d_v $, $ d_a $, and $ d_t $ are the dimensions of the visual, audio, and textual feature vectors, respectively. This multi-channel modal learning approach allows us to extract representations from each modality independently, without interference. Moreover, processing the modalities in parallel enhances computational efficiency.

\subsection{Personality Detection based on Multi-modal Analysis}

After extracting the necessary features from the aligned multi-modal data, we propose a comprehensive framework for personality prediction. This framework consists of two main stages: feature preprocessing and personality inference. In the first stage, we preprocess and align features from different modalities to ensure compatibility and enhance their representational capacity. In the second stage, we fuse these preprocessed features and employ a predictive model to infer personality traits. The overall architecture of the framework is depicted in Figure \ref{fig:framework}.

\subsubsection*{\ding{182}  \textbf{Time Frame Multi-Modality Data Reshaping} }
Due to the use of different embedding models for each modality, the extracted features have varying dimensions and formats. To address this inconsistency and capture temporal dependencies within each modality, we employ a Bidirectional Long Short-Term Memory (Bi-LSTM) network for each modality independently. Let $ X_i \in \{ v_{i,\text{face}}, v_{i,\text{background}}, v_{i,\text{audio}}, v_{i,\text{text}} \} $ represent the sequence of features for the $ i $-th modality of a video $ V $, where $ i $ indexes the modalities (face, background, audio, text). Each modality provides a sequence of feature vectors over time: $ X_i = \{ X_{i,1}, X_{i,2}, \dots, X_{i,n} \} $, where $ n $ is the total number of timestamps (standardized to 70 as previously described). Each modality's sequence $ X_i $ is processed through a Bi-LSTM network to capture temporal patterns and reshape the features into a consistent format:
\begin{equation}\label{eq5}
	{F}_{i,t}=T           \left(TS\left(\mathbf{X}_{i, {\tau}}\right), TS\left(\mathbf{X}_{i, {\tau}+1}\right), W_{i, t}\right)
\end{equation}
where $TS(\cdot)$ is a Bi-LSTM unit, $W_{i,t}$ is the weight of Bi-LSTM unit. $F_{i,t}$ is Bi-LSTM output of last layer. $T$ is the Bi-LSTM calculation of different frame modality data input. The output of Bi-LSTM takes different modality features into the same shape. Here, $ W_i $ represents the weights of the Bi-LSTM for modality $ i $, and $ F_i = \{ F_{i,1}, F_{i,2}, \dots, F_{i,n} \} $ is the output sequence of the Bi-LSTM for modality $ i $. This process ensures that the features from all modalities have compatible dimensions, facilitating subsequent fusion.

\subsubsection*{\ding{183}  \textbf{Temporal Fusion via Self-Attention Mechanism}}
To enhance the temporal representation within each modality, we apply a self-attention mechanism to the Bi-LSTM outputs. The self-attention mechanism allows the model to weigh the importance of different time steps, enabling it to focus on the most informative features for personality prediction. For each modality $ i $, we compute the self-attention output $ U_i $ as follows:
\begin{equation}\label{eq6}
        \mathbf{U}_{i}=\sum_{t=1}^{n} s_{i,t} \cdot \left(\mathbf{w}_{i,t} \mathbf{F}_{i,t}\right)
\end{equation}
where $\mathbf{w}_{i,t}$ is self-attention weight of one modality. $U_{i}$ is the self-attention output of one modality feature. $s_{i,t}$ is the self-attention score, which is calculated as follows: 
\begin{equation}\label{eq7}
s_{i, t}=\operatorname{softmax}\left(\frac{\mathbf{Q}_{i,t} \cdot\left(\mathbf{K}_{i,t}\right)^T}{\sqrt{d_k}}\right) \cdot \mathbf{V}_{i,t}
\end{equation}
where $d_k$ is used to normalize the value, queries $\mathbf{Q}_{i,t}=$ $\mathbf{W}_{i,t}^{Q} N\left(\mathbf{W}_k \mathbf{F}_i\right)$, keys $\mathbf{K}_{i,t}=\mathbf{W}_{i,t}^{K} N\left(\mathbf{W}_k \mathbf{F}_i\right)$, and values $\mathbf{V}_{i,t}=$ $\mathbf{W}_{i,t}^{V}N\left(\mathbf{W}_k \mathbf{F}_i\right)$. $N(\cdot)$ is the normalized function.

\subsubsection*{\ding{184}  \textbf{Cross-Modal Fusion with Bilinear Transformation} }
While the self-attention mechanism captures intra-modality temporal dependencies, integrating information across modalities provides complementary insights. To achieve cross-modal fusion, we employ a bilinear transformation to model the interactions between modalities. Let us denote the modalities as $ a, b, c, d $. For modality $ a $, we compute its interaction with the other modalities using bilinear pooling:
\begin{equation}
Z_i^a = \operatorname{MLP}\left( \left[ Z_{ab}, Z_{ac}, Z_{ad} \right] \right)
\end{equation}
where $ [ \cdot ] $ denotes concatenation, and $ Z_{ab}, Z_{ac}, Z_{ad} $ are the interaction features between modality $ a $ and modalities $ b, c, d $, respectively, computed as:
\begin{equation}
Z_{aj} = \sum_{t=1}^{n} \sigma\left( F_{i,t}^a W^{aj} \left( F_{i,t}^j \right)^\top \right), \quad \text{for } j \in \{ b, c, d \}
\end{equation}
In this equation, $ \sigma $ is an activation function such as ReLU, $ W^{aj} $ is a learnable weight matrix representing the interaction between modalities $ a $ and $ j $, and $ F_{i,t}^a $, $ F_{i,t}^j $ are the features of modalities $ a $ and $ j $ at time $ t $. This process captures the pairwise interactions between modality $ a $ and the other modalities, enriching modality $ a $'s representation with cross-modal information. The same procedure is applied to each modality to obtain fused representations $ Z_i^a, Z_i^b, Z_i^c, Z_i^d $.

\subsubsection*{\ding{185}  \textbf{Feature Aggregation and Personality Prediction} }
We then combine the self-attention outputs $ U_i $ and the cross-modal fused representations $ Z_i^i $ for each modality:
\begin{equation}
S_i = \operatorname{softmax}\left( \operatorname{concat}\left( U_i, Z_i^i \right) W_i \right)
\end{equation}
Here, $ W_i $ is a learnable weight matrix, and $ \operatorname{concat} $ denotes concatenation. This step integrates both the temporal and cross-modal information into a unified representation for each modality. Finally, we concatenate the embeddings $ S_i $ from all modalities to form the final embedding $ S $:
\begin{equation}
    S = \operatorname{concat}\left( S_a, S_b, S_c, S_d \right)
\end{equation}
The final embedding $ S $ encapsulates comprehensive information from all modalities. We feed $ S $ into a Multi-Layer Perceptron (MLP) to predict the personality traits:
\begin{equation}
    \mathbf{y} = [ y_1, y_2, y_3, y_4, y_5 ] = \operatorname{MLP}\left( S \right)
\end{equation}
where $ y_j $ represents the predicted score for the $ j $-th personality trait. The MLP serves as the classifier or regressor that maps the high-dimensional fused features to the personality trait predictions. In this way, we can effectively leverage both temporal dependencies and cross-modal interactions to enhance personality prediction from short videos. By processing each modality through a Bi-LSTM and self-attention mechanism, and then fusing them via bilinear transformations, the model captures rich and complementary information necessary for accurate personality inference.

\begin{figure}[t]
\centering
\includegraphics[width=0.45\textwidth]{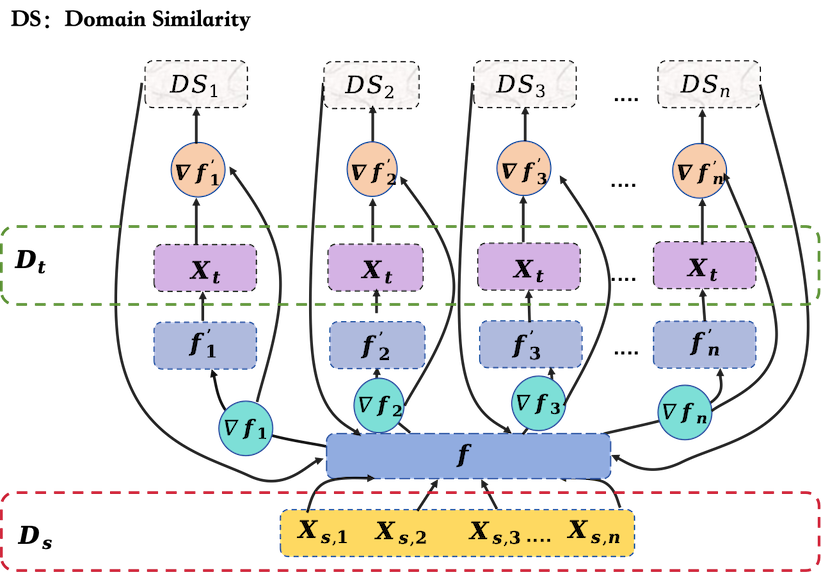}
\caption{The Flowchart of Domain Adaptation method.}
\label{fig:framework_ds}
\end{figure}

\subsection{Enhance Personality Analysis via Domain Adaptation}

Personality detection from online short videos presents significant challenges due to the diverse and dynamic nature of content across different platforms, cultures, and user demographics. Models trained on data from one domain often struggle to generalize to other domains because of variations in video styles, linguistic expressions, and contextual cues. Moreover, collecting extensive labeled data for every new target domain is often impractical. To address these challenges, we propose a domain adaptation method that enhances personality analysis by effectively leveraging data from multiple source domains to improve performance in a target domain with limited labeled examples.

\subsubsection*{\ding{182}  \textbf{Domain Adaptation for Few-Shot Learning}}
In our scenario, we have access to labeled data from $ k $ source domains, each providing a substantial amount of data for training. The target domain, however, has only a few labeled examples available. Our objective is to train a model $ f $ (as depicted in Figure \ref{fig:framework_ds}) that performs well on the target domain by utilizing the rich information from the source domains along with the limited data from the target domain. Here, the source domains are denoted by $ \mathcal{D}_s = \{ x_{s,1}, x_{s,2}, \dots, x_{s,k} \} $, where $ x_{s,i} $ represents the dataset from the $ i $-th source domain. The target domain is $ \mathcal{D}_t = x_t $, containing the few-shot data. Our aim is to find the optimal model parameters $ \theta $ that minimize the loss on the target domain:
\begin{equation}\label{eq13}
   \min_{\theta} \mathcal{L}(f(\theta), \mathcal{D}_t),
\end{equation}
where $ \mathcal{L} $ is the loss function, specifically the Mean Squared Error (MSE):
\begin{equation}\label{eq18}
\operatorname{MSE} = \frac{1}{N} \sum_{i=1}^N \left( \hat{y}_{ik} - y_{ik} \right)^2.
\end{equation}
Here, $ \hat{y}_{ik} $ is the predicted value for the $ k $-th dimension of the Big Five personality traits, and $ y_{ik} $ is the corresponding true value.

\subsubsection*{\ding{183}  \textbf{Proposed Method}}

To effectively adapt the model to the target domain, we introduce a gradient-based domain adaptation method that operates in two main stages: training on the source domains and adapting to the target domain using domain similarity measures.

In the first stage, we begin by initializing $ k $ models $ f_i $ with identical parameters $ \theta $, corresponding to each of the $ k $ source domains. Each model $ f_i $ is then trained on its respective source domain data $ x_{s,i} $, resulting in an updated model $ f'_i $. During this training, we compute the gradient of the loss function with respect to the model parameters for each source domain, denoted as $ \nabla f_i $:
\begin{equation}\label{eq14}
   \nabla f_i = \nabla_{\theta} \mathcal{L}\left( f(\theta), x_{s,i} \right).
\end{equation}
These gradients capture how the model parameters need to be adjusted to minimize the loss on each source domain, effectively encoding domain-specific learning directions.

In the second stage, we focus on adapting the model to the target domain by utilizing the few-shot examples available. We feed the target domain data $ x_t $ into each of the updated source models $ f'_i $ and compute the gradient of the loss function with respect to the model parameters:
\begin{equation}\label{eq15}
   \nabla f'_t = \nabla_{\theta} \mathcal{L}\left( f'_i(\theta), x_t \right),
\end{equation}
where $ \nabla f'_t $ represents the gradient for the target domain using the $ i $-th updated source model. We then measure the similarity between the target domain and each source domain by calculating the cosine similarity between their gradients:
\begin{equation}\label{eq16}
\text{sim}_{\text{cos}}(\nabla f'_t, \nabla f_i) = \frac{ \nabla f'_t \cdot \nabla f_i }{ \| \nabla f'_t \| \times \| \nabla f_i \| },
\end{equation}
where $ \cdot $ denotes the dot product, and $ \| \cdot \| $ denotes the Euclidean norm. The similarity scores $ s_i = \text{sim}_{\text{cos}}(\nabla f'_t, \nabla f_i) $ quantify how closely each source domain is related to the target domain in terms of gradient directions.

\begin{algorithm}[t]
\caption{Domain Adaptive Algorithm}\label{alg:alg1}
\begin{algorithmic}
\STATE 
\STATE {\textsc{Input}}$(Initial \,Model \,parameter\, \mathbf{\theta}$
\STATE\hspace{1.0cm}$\mathbf{k} \,Source\, Domain \,dataset \,\mathbf{X_{si}}$
\STATE\hspace{1.0cm}$few\,shot\,data\,\mathbf{X_t}\,in\,Target\,Domain$
\STATE\hspace{1.0cm}$number \,of\,iterations\,\mathbf{N})$

\STATE$\textbf{for inter} \subset \textbf{range N}$
\STATE\hspace{0.5cm}$ \textbf{for i} \subset \textbf{range k}$
\STATE\hspace{0.9cm}$Update \,source\, domain\, model \,parameter$
\STATE\hspace{1.2cm}$with \,\mathbf{X_{si}}$
\STATE\hspace{0.9cm}$Compute \,Source \,Domain\, Gradient$
\STATE\hspace{0.9cm}$by \, Equation \,(12)$
\STATE\hspace{0.9cm}$Compute \,Target \,Domain\, Gradient$
\STATE\hspace{1.2cm}$by \, Equation \,(13)$
\STATE\hspace{0.9cm}$Calculate \,Domain\, Similarity$
\STATE\hspace{1.2cm}$by \, Equation \,(14)$
\STATE\hspace{0.5cm}$\textbf{end}$
\STATE\hspace{0.9cm}$Update \,initial\, model \,parameter\, \mathbf{\theta}$
\STATE\hspace{0.9cm}$ with \,Equation\,(15)$
\STATE $\textbf{end}$
\end{algorithmic}
\label{alg1}
\end{algorithm}

Using these similarity scores, we update the initial model parameters $ \theta $ by weighting the gradients from each source domain according to their relevance to the target domain:
\begin{equation}\label{eq17}
\theta \leftarrow \theta - \alpha \sum_{i=1}^k s_i \cdot \nabla_{\theta} \mathcal{L}\left( f'_i(\theta), x_t \right),
\end{equation}
Where $ \alpha $ is the learning rate. This update step ensures that the model emphasizes learning from source domains that are more similar to the target domain, thereby improving its ability to generalize despite the limited data available in the target domain.

By integrating the gradient information from both the source and target domains, the model effectively transfers knowledge from relevant source domains while adapting to the specific characteristics of the target domain. This approach addresses the challenges associated with domain shift and few-shot learning in personality detection from online short videos.

\subsubsection*{\ding{184}  \textbf{Training Procedure}}

To further enhance the model's performance, we incorporate an adaptive learning rate strategy as shown in Algorithm \ref{alg:alg1}. The overall training procedure begins with the initialization of the model parameters $ \theta $. During each iteration of training, the following steps are executed:

First, each model $ f_i $ is trained on its respective source domain data $ x_{s,i} $, resulting in updated models $ f'_i $. The source domain gradients $ \nabla f_i $ are computed using Equation \eqref{eq14}. These gradients reflect how each source domain influences the model parameters.

Next, the adaptation to the target domain is performed by feeding the few-shot target examples $ x_t $ into each of the updated source models $ f'_i $ and computing the target domain gradients $ \nabla f'_t $ as in Equation \eqref{eq15}. The domain similarity scores $ s_i $ are then calculated using Equation \eqref{eq16}, providing a measure of how relevant each source domain is to the target domain based on the alignment of their gradient directions.

Finally, the model parameters $ \theta $ are updated using the weighted gradients from the source domains as specified in Equation \eqref{eq17}. The learning rate $ \alpha $ may be adjusted adaptively during training to improve convergence and generalization, following strategies from prior work.

This iterative process continues until the model converges or a stopping criterion is met. By continually adjusting the model parameters based on domain similarity, the method ensures that the model learns from the most relevant source domains while adapting to the unique aspects of the target domain. By using gradient-based domain similarity, the model selectively integrates knowledge from source domains that are most similar to the target domain, thereby enhancing its performance in few-shot learning scenarios. This method is particularly valuable for personality detection in online short videos, where data diversity and scarcity are significant challenges.

\section{Experimental Analysis}\label{sec:experiment}

\subsection{Data Preparation}

In our work, we utilize the First Impressions dataset created by Biel and Gatica-Perez from the 2016 ChaLearn competition \cite{ponce2016chalearn} for training and evaluation. This dataset comprises 10,000 short video clips, each lasting 15 seconds. All videos were collected from the online social platform YouTube and feature individuals in everyday interview scenarios. The videos are labeled with scores for the Big Five personality traits, indicating each interviewee's personality across these five dimensions. The personality trait scores are represented within the range $[0, 1]$. Additionally, text transcriptions for each video are provided.

To implement our domain adaptation training method with the First Impressions dataset, we employ the K-means clustering algorithm to group the entire dataset into 20 topics based on their textual features. Compared to other modalities, textual data is usually closely related to the topics discussed in the videos. Manually labeling the topic of each video is impractical; therefore, clustering offers an efficient and effective way to group videos with similar content by leveraging their contextual features. We assign domain labels numbered from 1 to 20. Subsequently, our multi-modal feature extraction framework is applied to process the videos in each domain separately.

\subsection{Baseline Methods}

To evaluate the effectiveness of our method under few-shot learning conditions, we compare its performance with the following baseline methods:

\begin{itemize}
    \item \textbf{IVL} \cite{gorbova2018integrating}: This method employs LSTM cells to process video audio, text transcriptions, and the speaker's facial features.
    \item \textbf{DCC} \cite{guccluturk2016deep}: This approach introduces an end-to-end video understanding residual neural network, utilizing video and audio data to analyze the speaker's personality.
    \item \textbf{Evolgen} \cite{subramaniam2016bi}: This method uses audio clips from the videos and employs an LSTM network to predict personality traits.
    \item \textbf{ICC} \cite{sun2018personality}: In this method, CNN and LSTM architectures are used to extract textual features, which are then utilized for personality prediction.
    \item \textbf{DBR} \cite{zhang2016deep}: This method processes facial image data and audio data from the videos using convolutional neural networks. The two modalities are then fused for the personality regression task.
    \item \textbf{Fine-tune}: In this method, an MLP model is pre-trained on the source domain data and then fine-tuned using few-shot data from the target domain.
\end{itemize}

\begin{table*}[t]
\centering
\caption{Comparison of accuracy (\%) on Big Five personality traits across different models. The best results are highlighted in bold. Our proposed model achieves superior performance in terms of average accuracy and individual personality traits.}
\label{tab:results}
\resizebox{0.98\textwidth}{!}{%
\begin{tabular}{lcccccc}
\toprule
\textbf{Model} & \textbf{Average} & \textbf{Extraversion} & \textbf{Neuroticism} & \textbf{Agreeableness} & \textbf{Conscientiousness} & \textbf{Openness} \\
\midrule
IVL        & $63.87_{\pm 2.34}$  & $68.27_{\pm 4.94}$        & $64.42_{\pm 4.07}$      & $61.67_{\pm 4.19}$        & $65.77_{\pm 3.96}$           & $59.24_{\pm 3.70}$  \\
DCC        & $59.32_{\pm 6.06}$   & $51.76_{\pm 7.37}$       & $52.79_{\pm 8.87}$       & $54.41_{\pm 8.16}$        & $51.57_{\pm 7.87}$             & $56.62_{\pm 8.81}$    \\
Evolgen        & $59.93_{\pm 3.70}$   & $58.43_{\pm 2.71}$       & $60.48_{\pm 3.50}$       & $56.73_{\pm 7.42}$        & $61.30_{\pm 3.75}$             & $59.68_{\pm 5.41}$    \\
ICC        & $67.30_{\pm 1.66}$   & $68.27_{\pm 1.04}$       & $67.24_{\pm 2.34}$       & $66.56_{\pm 2.03}$        & $68.15_{\pm 2.13}$             & $66.26_{\pm 2.85}$    \\
DBR   & $65.19_{\pm 1.25}$   & $64.80_{\pm 2.32}$       & $65.15_{\pm 1.91}$       & $65.69_{\pm 1.68}$         & $63.84_{\pm 1.43}$             & $67.43_{\pm 1.96}$    \\
Fine-tune   & $77.78_{\pm 0.49}$   & $77.40_{\pm 0.71}$       & $77.26_{\pm 0.58}$       & $79.19_{\pm 0.75}$         & $77.08_{\pm 0.67}$             & $77.97_{\pm 0.66}$    \\
\rowcolor{gray!20}
\textbf{Ours}   & $\mathbf{77.92}_{\pm 0.52}$  & $\mathbf{77.54}_{\pm 0.72}$        & $\mathbf{77.42}_{\pm 0.48}$       & $\mathbf{79.33}_{\pm 0.67}$         & $\mathbf{77.26}_{\pm 0.71}$             & $\mathbf{78.08}_{\pm 0.76}$    \\
\bottomrule
\end{tabular}
}
\end{table*}

\subsection{Experimental Settings}

All experiments are conducted on a system running Ubuntu with an NVIDIA GeForce RTX 3090 GPU. We train our model using 10 few-shot data samples from the target domain. The remaining data in the target domain are split into validation and test sets with a ratio of 2:8. The initial learning rate $\alpha$ is selected through grid search as a hyperparameter. We employ the AdamW optimizer to train the model.

\subsection{Evaluation Metric}

For the evaluation, we select each of the 20 domains as the target domain in turn, using all remaining domains as source domains. During the domain adaptation training process, for each training epoch, we randomly select 10 few-shot data samples from the target domain to compute domain similarity and use the remaining data in the target domain for validation and testing. To assess the performance of our method, we use the average accuracy of each individual personality trait to compare different models. The average accuracy is defined as:
\begin{equation}\label{eq19}
\operatorname{Accuracy} = \frac{1}{N} \sum_{i=1}^N \left( 1 - \left| \hat{y}_{ik} - y_{ik} \right| \right) \times 100\%,
\end{equation}
where $N$ is the number of samples, $\hat{y}_{ik}$ is the predicted value, and $y_{ik}$ is the true value for the $k$-th personality trait of the $i$-th sample.

\begin{table*}[t]
\centering
\caption{Effectiveness of Different Components in Our Framework. The table compares the average accuracy (\%) and individual personality trait accuracies when certain components are removed or different optimizers are used. The best results are highlighted in bold.}
\label{tab:variant_results}
\resizebox{0.98\textwidth}{!}{%
\begin{tabular}{lcccccc}
\toprule
\textbf{Variants} & \textbf{Average} & \textbf{Openness} & \textbf{Extraversion} & \textbf{Agreeableness} & \textbf{Neuroticism} & \textbf{Conscientiousness} \\
\midrule
w/o Domain Similarity       & 77.20   & 77.09    & 77.52        & 78.63         & 77.59       & 77.14             \\
w/o Adaptive Learning Rate  & 77.38   & 77.70    & 77.07        & 77.96         & 77.19       & 77.04             \\
DAL-SGD                     & 77.32   & 77.57    & 77.23        & 78.45         & 76.84       & 76.49             \\
DAL-Adam                    & 77.22   & 77.08    & 77.21        & 77.95         & 77.27       & 76.60             \\
\rowcolor{gray!20}
\textbf{DAL-AdamW (Ours)}   & \textbf{77.92}   & \textbf{78.08}    & \textbf{77.54}        & \textbf{79.33}         & \textbf{77.42}       & \textbf{78.08}             \\
\bottomrule
\end{tabular}
}
\end{table*}

\subsection{Overall Performance Evaluation}
The results in Table \ref{tab:results} demonstrate that our proposed model achieves superior accuracy in predicting the Big Five personality traits from short video data. With an average accuracy of \textbf{77.92\%}, our model outperforms all baseline methods, indicating its effectiveness in capturing personality cues across multiple modalities. Compared to the closest baseline, the Fine-tune method with an average accuracy of 77.78\%, our model shows consistent improvements across all personality traits. Specifically, we achieve higher accuracies in Extraversion (77.54\% vs. 77.40\%), Neuroticism (77.42\% vs. 77.26\%), Agreeableness (79.33\% vs. 79.19\%), Conscientiousness (77.26\% vs. 77.08\%), and Openness (78.08\% vs. 77.97\%). These gains, though modest, reflect the robustness of our approach to enhancing personality prediction.

Our model significantly outperforms other baselines such as IVL, DCC, Evolgen, ICC, and DBR, which have average accuracies ranging from approximately 59\% to 67\%. The substantial margin, over 10 percentage points, between our model and these methods underscores the effectiveness of our techniques. This improvement can be attributed to our timestamp-based modality alignment, which ensures precise synchronization of multi-modal data, and the use of self-attention mechanisms and cross-modal fusion that enhance feature representation. Furthermore, the incorporation of gradient-based domain adaptation allows our model to generalize well to target domains with limited labeled data. By leveraging similarities between source and target domains, our method efficiently utilizes few-shot data, addressing common challenges in personality analysis tasks. In summary, the superior performance of our model across all five personality traits validates the effectiveness of our multi-modal framework. The consistent accuracy improvements highlight our model's ability to capture complex behavioral cues and adapt to diverse data domains, offering a significant advancement over existing methods in personality prediction from short videos.

\begin{figure}[t]
\centering
\includegraphics[width=0.45\textwidth]{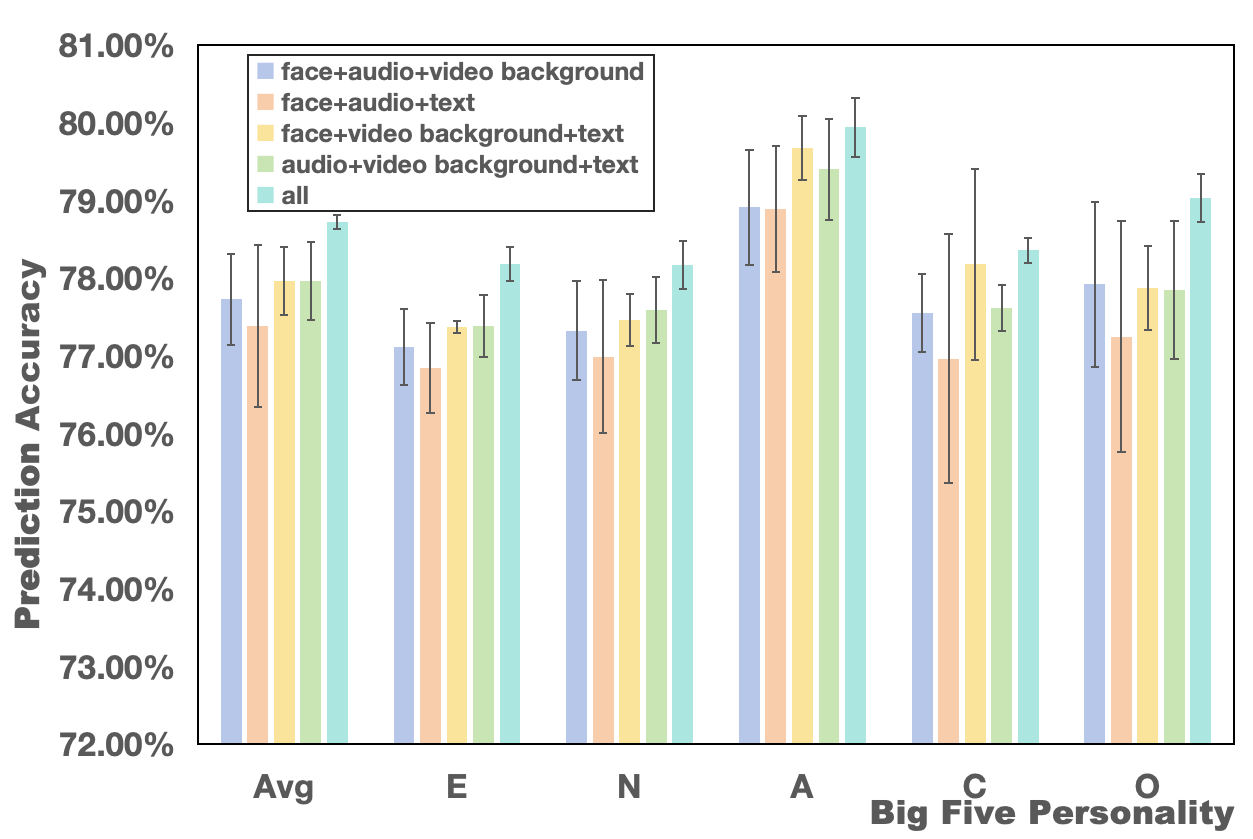}
\caption{Multi-modal Sensitivity}
\label{fig:sub2-1}
\end{figure}

\begin{figure}[t]
\centering
\includegraphics[width=0.45\textwidth]{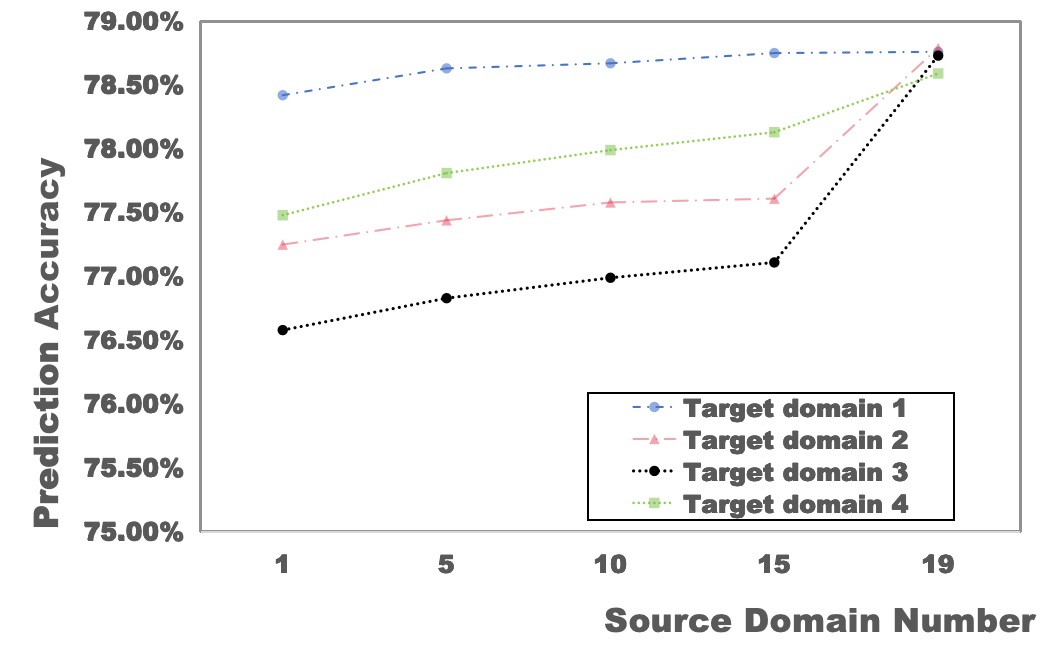}
\caption{Multimodal Adaptation Effectiveness Evaluation}
\label{fig:results_different_source_number}
\end{figure}

\subsection{Multi-modal Adaptation Effectiveness}
To see how different modality combinations work with our framework, we remove one modality from the original four modalities and get four group modality combinations. Then, training different modalities in combination with the proposed domain adaptive learning method, the average values of single personality prediction accuracy are plotted in Fig. \ref{fig:sub2-1}. From Fig. \ref{fig:sub2-1}, we can see that the adaptation method with face, audio, text, and video background four modalities shows the best performance compared with the other four modalities combinations with only three modalities. Fig. \ref{fig:sub2-1} also shows the standard deviation for each modalities group. Our method with all four modalities shows a minimum standard deviation, while the model trained with only three modalities (facial image, speaker audio, text) has the maximum standard deviation. This means that compared with a method with fewer modalities, our method with more modalities shows less wave on the model performance. Our method has a more stable personality prediction performance.

\subsection{The Number of Source Domain Test}
For our Domain Adaptive learning method, the number of source domains is a hyper-parameter. At first, we take all remaining data as Source Domains. However, for performance evaluation, we can see that different source domains may have an impact on the performance of our domain adaptive method. Thus, we change the source domain numbers to 1, 5, 10, and 15, respectively and randomly select the corresponding number of source domains to train our model. The experiment results are summarized in Fig. \ref{fig:results_different_source_number}, which plots the prediction accuracy of different target domains 1, 2, 3, and 4 with different source domain numbers. From the result, it can be concluded that the model performance improves with the increase of the Source Domain number. However, it can also be noticed that even with the same number of source domains, the model performance shows great difference, as they select different domains as source domains.

\subsection{The Effectiveness Of Principal Components}
The results presented in Table \ref{tab:variant_results} demonstrate the significant impact of the key components in our framework on the accuracy of personality trait predictions. When the domain similarity calculation is removed, there is a noticeable decrease in average accuracy from 77.92\% to 77.20\%, marking a drop of 0.72\%. This reduction is consistent across all individual personality traits, with Openness decreasing from 78.08\% to 77.09\% and Agreeableness falling from 79.33\% to 78.63\%. These findings highlight the crucial role of the domain similarity calculation in our gradient-based domain adaptation method. By quantifying the relevance of source domains to the target domain, this component enables the model to prioritize learning from the most pertinent data, thereby enhancing generalization and performance, especially in few-shot learning scenarios.

Excluding the adaptive learning rate also affects the model's performance, resulting in an average accuracy of 77.38\%, which is a reduction of 0.54\% compared to the full model. The most significant decrease is observed in Conscientiousness, dropping from 78.08\% to 77.04\%. Other personality traits exhibit slight declines as well. This suggests that the adaptive learning rate contributes to more efficient and effective optimization by adjusting the step size during training. It helps the model to converge better and avoid local minima, leading to improved accuracy across various personality traits. We also test model performance under different optimizer choices. The model with SGD and Adam shows 0.6\%, 0.7\% performance decrease compared with AdamW optimizer.

\subsection{Domain Similarity Analysis}
\begin{figure}[t]
\centering
\includegraphics[width=0.45\textwidth]{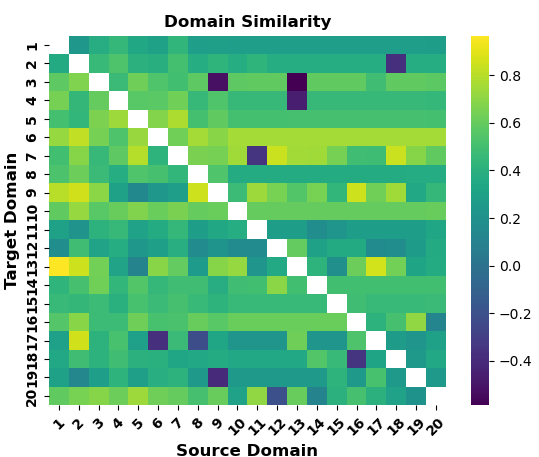}
\caption{Domain Similarity}
\label{fig:domain_similarity}
\end{figure}

In our method, we use the gradient to calculate the domain similarity, which can help the model work better in the target domain. Thus, the domain similarity calculation result can be used to measure the similarity of different domains. To evaluate how our method learns the similarity between different domains, we collect the final domain similarity calculation results, after training the model. We visualize the results by plotting a heat map, which can show the similarity result between two domains as shown in Figure \ref{fig:domain_similarity}. It can be seen that some domains have high similarity with others. For example, domain 13 has a high similarity with domain 1, and domain 17 has a high similarity with domain 13, etc. However, we can also notice that some domains have negative similarities with each other. For example, domain 3 and domain 4 show negative similarity with domain 13.

\section{Conclusion}\label{sec:conclusion}
In this paper, we investigate a multi-domain adaptation learning method with multi-modalities personality prediction. To improve model performance, we propose a timestamp modal alignment method to preprocess four different modalities in the short video. More importantly, we also propose a multi-domain adaptation method to improve model performance on few-shot data. We conduct experiments to evaluate model performance by comparing it with state-of-the-art baselines. 
% We also explore more on some important influence factors for proposed method. All the result shows that our proposed method demonstrates better performance on the few-shot case. However, there are still more space to explore and further improve prediction accuracy with few-shot data.
Future directions may include LLMs findings \cite{shu2024llm}, social network knowledge \cite{sun2023all} with sociological analysis \cite{sun2023self,cui2023event}, etc.

\normalem
\bibliographystyle{IEEEtran}
\bibliography{reference}

\begin{IEEEbiography}[{\includegraphics[width=1in,height=1in,clip,keepaspectratio]{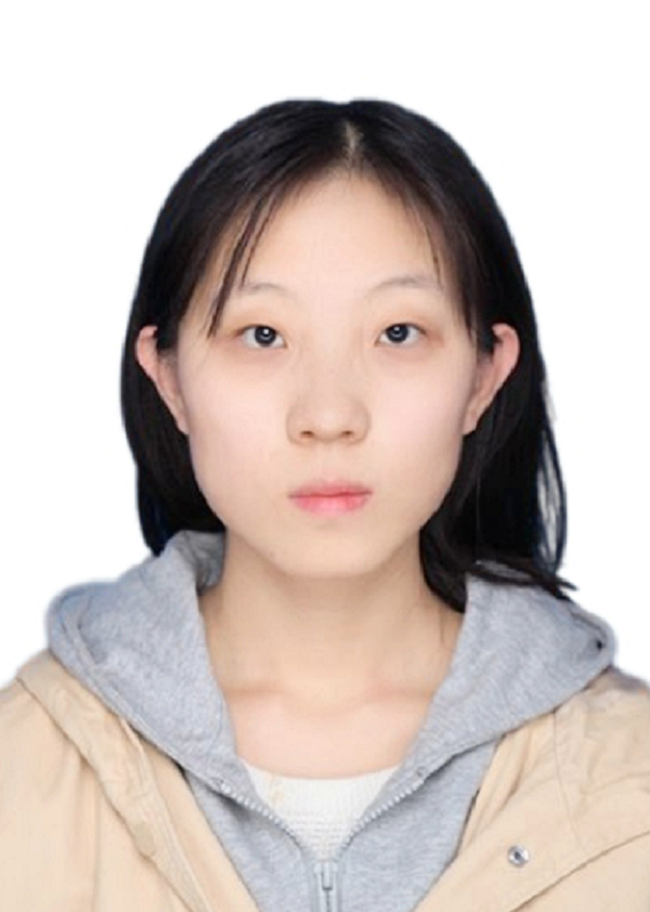}}]{Sixu An} is currently a Ph.D. student of the University Research Facility of Data Science and Artificial Intelligence (UDSAI) and Centre for Learning, Teaching and Technology (LTTC) in The Education University of Hong Kong. Before that, she received the M.Eng.degree from University College London in 2021 and the Bachelor degree from Bei Jing Jiao Tong University in 2020. Her research interest includes multi-modal video detection, Intelligent Tutoring System, etc.
\end{IEEEbiography}
\vspace{-10 mm}

\begin{IEEEbiography}[{\includegraphics[width=1in,height=1in,clip,keepaspectratio]{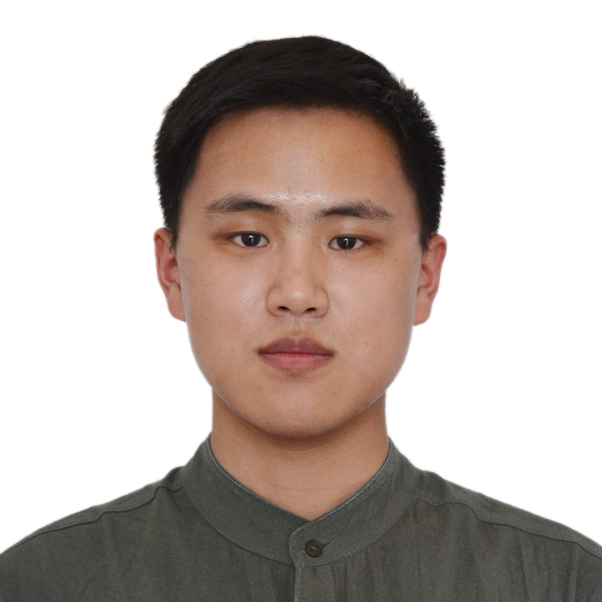}}]{Xiangguo Sun}  is a postdoctoral research fellow at The Chinese University of Hong Kong. He studied at  Zhejiang Lab as a visiting researcher in 2022. In the same year, he received his Ph.D. from Southeast University and won the Distinguished Ph.D. Dissertation Award. During his Ph.D. study, he worked as a research intern at Microsoft Research Asia (MSRA) and won the ''Award of Excellence''. He studied as a joint Ph.D. student at The University of Queensland hosted by ARC Future Fellow Prof. Hongzhi Yin from Sep 2019 to Sep 2021. His research interests include social computing and network learning. He has published 18 CORE A*, 15 CCF A, and 17 SCI (including 8 IEEE Trans), some of which appear in SIGKDD, ICLR, VLDB, The Web Conference (WWW), TKDE, TOIS, WSDM, TNNLS, CIKM, etc. He has 2 first-authored papers recognized as the Most Influential Papers in KDD 2023 and WSDM 2021. He was the winner of the Best Research Paper Award at KDD'23, which was the first time for Hong Kong and Mainland China. He was recognized as the "Social Computing Rising Star" in 2023 by CAAI.
\end{IEEEbiography}
\vspace{-10 mm}

\begin{IEEEbiography}[{\includegraphics[width=1in,height=1.25in,clip,keepaspectratio]{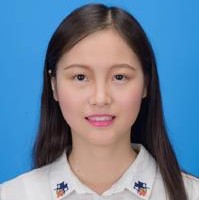}}]{Yicong Li}
is currently a Ph.D. student of the Data Science and Machine Intelligence (DSMI) Lab of Advanced Analytics Institute, University of Technology Sydney. Her research interests mainly focus on data science, graph neural networks, recommender systems, natural language processing and so on. In particular, her current research is focusing on explainable machine learning, especially the application in the recommendation area. She has published papers in international conferences and journals, such as KDD, TKDE, WSDM, CIKM, KSEM and IEEE Access. She has also reviewed papers in many top-tier conferences and journals, like AAAI, KDD, WWW, IJCAI, WSDM, ICONIP and so on. In addition, she has been invited to review manuscripts in IEEE Transactions on Neural Networks and Learning Systems (TNNLS), which is a top-tier journal in artificial intelligence. 
\end{IEEEbiography}

\vspace{-10 mm}

\begin{IEEEbiography}[{\includegraphics[width=1in,height=1in,clip,keepaspectratio]{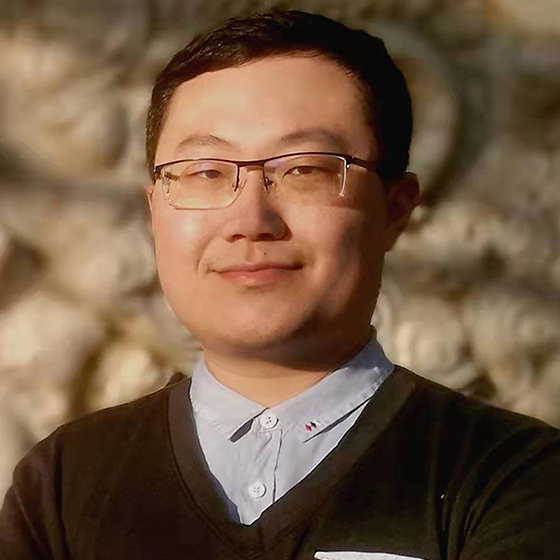}}]{Yu Yang}
is currently an Assistant Professor with the Centre for Learning, Teaching, and Technology (LTTC) at the Education University of Hong Kong. He received the M.Eng. degree in Pattern Recognition and Intelligence System from Shenzhen University in 2015 and the Ph.D. degree in Computer Science from The Hong Kong Polytechnic University in 2021. His research interests include spatiotemporal data analysis, representation learning on dynamic graphs, urban computing, and learning analytics.
\end{IEEEbiography}

\vspace{-10 mm}

\begin{IEEEbiography}[{\includegraphics[width=1in,height=1.25in,clip,keepaspectratio]{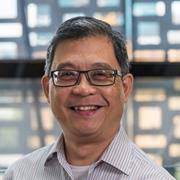}}]{Guandong Xu}
is a Chair Professor of Artificial Intelligence, Director of the University Research Facility of Data Science and Artificial Intelligence, and Director of the Centre for Learning, Teaching and Technology at The Education University of Hong Kong. He has published more than 240 papers in leading international journals and revered conference proceedings. His work encompasses a wide spectrum of fields, including recommender systems, information retrieval, data-driven knowledge discovery, and social computing, and continues to garner increasing citations from the academic community. He was listed in the top 2\% in the Stanford list of the world’s most-cited scientists for consecutive years.
He also serves as the founding Editor-in-Chief of the Human-centric Intelligent Systems Journal (Springer) and the Assistant Editor-in-Chief of the World Wide Web Journal (Springer). Moreover, he was elected as a Fellow of the Institution of Engineering and Technology (IET) in the United Kingdom and Fellow of the Australian Computer Society (ACS) in both 2021 and 2022. 
\end{IEEEbiography}

\vfill
\end{document}